\documentclass[preprint]{ptephy_v1}

\preprintnumber{XXXX-XXXX} 
\usepackage{hyperref}

\begin{document}

\title{Duality analysis in a symmetric group and its application to random tensor network models}


\author{Masayuki Ohzeki}
\affil{Graduate School of Information Sciences, Tohoku University, Miyagi 980-8564, Japan \\
Department of Physics, Tokyo Institute of Technology, Tokyo, 152-8551, Japan \\
Sigma-i Co., Ltd., Tokyo, 108-0075, Japan
\email{mohzeki@tohoku.ac.jp}}


\begin{abstract}%
The Ising model is the simplest model for describing many-body effects in classical statistical mechanics.
A duality analysis leads to its critical point under several assumptions.
The Ising model has $Z_2$-symmetry.
The basis of duality analysis is a nontrivial relationship between low- and high-temperature expansions.
However, discrete Fourier transformation automatically determines hidden the relationship.
The duality analysis can naturally extend to systems with various degrees of freedom, $Z_q$ symmetry, and random spin systems.
Furthermore, in the present study, we obtained the duality relation in a series of permutation models by considering the symmetric group $S_q$ and its Fourier transformation.
The permutation model in a symmetric group is closely related to random quantum circuits and random tensor network models, which are frequently discussed in the context of quantum computing and the holographic principle, a property of string theories and quantum gravity.
We provide a systematic approach using duality analysis to examine the phase transition in these models.
\end{abstract}

\subjectindex{A-41}

\maketitle

{\it Introduction:}
The Ising model is a well-known statistical-mechanical model that exhibits critical behavior in finite dimensions.
A nontrivial analysis of the Ising model in a two-dimensional square lattice, in which the Ising model has a single critical point, was initiated using the Kramers-Wannier duality \cite{Kramers1941}.
In general, the identification of the location of the critical point is nontrivial. 
It can be obtained by diagonalizing the transfer matrix on two-dimensional lattices and through numerical computation methods such as the transfer matrix and Markov-chain Monte Carlo methods.
However, a duality analysis greatly simplifies the estimation of the locations of critical points. 
The essence of duality analysis is to find a connection between the original and dual models.

Duality analysis can be generalized to cases with $Z_q$ symmetry using a Fourier transformation, as in the literature \cite{WuWang1976}.
This generalization is helpful for many-component spin models, such as the Potts, Villain \cite{Kogut1979}, and even the spin-glass models via the replica method \cite{Nishimori1979, Nishimori2002, Maillard2003, Nishimori2006, Ohzeki2008hl, Ohzeki2009, Ohzeki2011slope,Ohzeki2015}.
This technique is useful for analyzing the theoretical limitations of error-correcting codes, such as the toric \cite{Dennis2002, Ohzeki2012duality, Ohzeki2012} and color codes, and its depolarizing channel \cite{Bombin2014}.
Its generalization to ground states has also been discussed \cite{Ohzeki2013, Ohzeki2018gs,Miyazaki2020}.
In several models, such as the Ising model, the dual model is the same as the original model but with different parameters. 
These are called self-dual models.
However, most of these models are not necessarily self-dual.
In such cases, a duality analysis sometimes leads to the precise locations of critical points.
Remarkable cases have been observed in random spin systems.
Furthermore, systematic partial summation as a real-space renormalization, combined with a duality analysis, can improve the precision of the estimation of critical point locations \cite{Ohzeki2009, Ohzeki2015}. 
To generalize highly nontrivial cases, they simplified duality analyses into essential computation by using only a single equation to connect the original and its dual models.
An equality consists of the trivial edge Boltzmann factor for the original model and its dual models.
Therefore, applying a simplified duality analysis to nontrivial cases is straightforward, even for random spin systems.

In recent studies, statistical models with symmetric groups have attracted attention in quantum computation and holographic principles in string theory and quantum gravity.
In several types of quantum random circuits, the connection between entanglement entropy and free energy of a statistical model with a symmetric group has been established \cite{Zhou2019,Romain2019,Yimu2020,jian2020}.
In addition, a random tensor network model, a simple theoretical model for investigating holographic principles, was connected to the statistical model in a symmetric group.
For these cases, we developed a permutation model for the symmetric group.
This model has a long history \cite{Drouffe1979}.
A duality analysis was also discussed \cite{Drouffe1978}.
The Kramers and Wannier duality analysis was generalized to the case of non-Abelian groups with modern applications, as in literature \cite{Buchstaber2003}.
However, generalizing the estimation of critical points using duality transformation is not straightforward for the non-self dual model.
In the present study, we used Fourier transformation in the symmetric group and established a single equation to estimate the location of the critical point in a random tensor network model.

{\it Duality in the $Z_q$ model:}
First, we review the duality analysis of spin models with $Z_q$ symmetry.
The Fourier transformation of the function $f(\phi)$ in $Z_q$ symmetry is given by
\begin{equation}
    \hat{f}(\lambda) = \frac{1}{\sqrt{q}} \sum_{\phi=0}^{q} f(\phi) \rho(\phi)^{\lambda},
\end{equation}
where $\rho(\phi) = \exp(i\phi)$.
Note that $\lambda$ is an exponent.
This well-known orthogonality is satisfied by $\sum_{\phi} \rho(\phi)^{\lambda}\rho(\phi)^{\lambda'} = q \delta_{\lambda,\lambda'}$.
In addition, $\rho(\phi_1)^{\lambda}\rho(\phi_2)^{\lambda} = \rho(\phi_1 + \phi_2)^{\lambda}$.
Here, we provide a partition function of a spin model, which includes the Potts model as a special case and the spin-glass model handled by the replica method, as follows:
\begin{eqnarray}
    Z = \sum_{\{\phi\}} \prod_{(ij) \in E} f(\phi_i - \phi_j).
\end{eqnarray}
For simplicity, we set the model as a square lattice.
Edge $(i,j) \in E$ is a set on each bond in the square lattice, and the edge Boltzmann factor $f(\phi)$ is set at each bond.
Here, we use another expression based on the difference between the sites in the square lattice: $\phi_{ij} = \phi_i - \phi_j$.
The difference around each face $\square_k$ in the square lattice must satisfy 
\begin{eqnarray}
    \sum_{(ij) \in \square_k } \phi_{ij} = 0~({\rm mod}~q).
\end{eqnarray}
where $\sum_{(ij) \in \square_k }f(\phi_{ij}) = f(\phi_{12}) + f(\phi_{23}) + f(\phi_{34}) + f(\phi_{41})$.
For example, we take a unit plaquette on the square lattice.
The top-left site is denoted by $1$, and we set clockwise the sites denoted by $2$ (top-right), $3$ (bottom-right), and $4$ (bottom-left), respectively.
This enables us to rewrite the partition function as
\begin{eqnarray}
    Z = q \sum_{\{\phi\}} \prod_{(ij) \in E} f(\phi_{ij}) \prod_{k} \delta_q \left(\sum_{(ij) \in \square_k } \phi_{ij} \right).
\end{eqnarray}
Here, summation is performed for all possible combinations of $\phi_{ij} = 0,1,\ldots,q-1$.
The coefficient $q$ arises from the arbitrariness of the summation over $\phi_i$, but it is negligible within thermodynamic limits.
The Kronecker delta-like function $\delta_q(x)$ is defined as follows:
\begin{eqnarray}
    \delta_q(\phi) = \left\{
    \begin{array}{ll}
    1 & (\phi \equiv 0 \mod q) \\
    0 & {\rm otherwise}
    \end{array}
    \right.
\end{eqnarray}
Here, we use the following identity: 
\begin{eqnarray}
\delta_q(\phi) = \frac{1}{q}\sum^{q-1}_{\lambda=0} \rho(\phi)^{\lambda}.
\end{eqnarray}
The partition function can then be rewritten as
\begin{eqnarray}
    Z = \frac{q}{q^N} \sum_{\{\lambda\}}\prod_{(kk') \in E} \sum_{\phi_{ij}}f(\phi_{ij})\rho(\phi_{ij})^{\lambda_k}\rho(-\phi_{ij})^{\lambda_{k'}},
\end{eqnarray}
where $N$ is the number of sites, and
$(kk')$ is the edge perpendicular to the original bond $(ij)$.
The definition of the Fourier transformation yields
\begin{eqnarray}
    Z = q \sum_{\{\lambda\}}\prod_{(kk') \in E} \hat{f}(\lambda_k - \lambda_{k'}).
\end{eqnarray}
This is the dual expression of the spin model with a $Z_q$ symmetry.
This derivation is identical to that presented in \cite{WuWang1976}.

The critical point, under the assumption of a unique phase transition under a certain condition, satisfies
\begin{eqnarray}
    f(0) = \hat{f}(0), \label{con1}
\end{eqnarray}
where $0$ denotes a trivial state (parallel spin) of the edge Boltzmann factor.
In addition, $\hat{f}(0)$ is the summation of all possible states in the original edge Boltzmann factor.
This equation yields the well-known critical points of the Ising model, $\exp(-2K)=\sqrt{2}-1$ from $f(0)=\exp(K)$ and $\hat{f}(0)=(\exp(K)+\exp(-K))/\sqrt{2}$$K$, and those of the Potts model, $\exp(K)=\sqrt{q}+1$ from $f(0)=\exp(K)$ and $\hat{f}(0)=(\exp(K)+q-1)/\sqrt{q}$  Here, $K$ is the coupling constant.
This equation is also available for the spin-glass model.
For instance, in the case of the $\pm J$ Ising model, where the concentrations of the ferromagnetic and antiferromagnetic interactions are $p$ and $1-p$, respectively, we obtain $-p\log p -(1-p)\log(1-p) = 1/2$.
The equality estimated the location of the multi-critical point on the Nishimori line, which is a special subspace stemming from gauge symmetry, to be approximately $p_c=0.889972$ \cite{Nishimori1979,Nishimori2002,Maillard2003}.
However, the dual model differed from the original one in this case.
Hence, a duality analysis does not necessarily estimate the exact locations of the critical points.
Systematic summation in the square lattice improves the precision of the estimation and duality analysis \cite{Ohzeki2008hl,Ohzeki2009, Ohzeki2011slope, Ohzeki2015}.
Thus, Equation (\ref{con1}) is an approximate estimator.
A duality analysis in a random spin system is performed using the replica method, where the power of the partition function, namely, the replicated system, is considered.
Subsequently, we use the natural number $n$ as the index of the replicated system.
The analytical continuation $n \to 0$ predicts the original quenched random-spin system results.
The duality analysis leads to exact solutions for $n=1,2$.
The case of $n=3$ is also discussed, and the precise locations of the critical points are estimated.
Summarizing these speculations, we expect the result when $n \to 0$ is not far from the exact solution.

{\it Permutation model:}
In this study, we introduced a permutation model.
The simplest model is given by 
\begin{eqnarray}
    E(\boldsymbol{\sigma}) = - J \sum_{(ij) \in E} C(\sigma_i\sigma_j^{-1}), \label{rtn}
\end{eqnarray}
where $\sigma_i \in S_q$ is a permutation that characterizes the degrees of freedom of the model.
Here, $S_q$ denotes the symmetry group constituting permutations of an integer $q$.
In addition, $C(\sigma)$ is a cycle-counting function that counts the number of cycles in the permutation $\sigma$.
For instance, $\sigma_i(1) = 2$, $\sigma_i(2)=1$, $\sigma_i(3) =3$ $\sigma_j(1) = 3$, $\sigma_j(2)=2$, and $\sigma_j(3) =1$.
Assume that $\sigma = \sigma_i\sigma_j^{-1}$
Then, $\sigma(1) = 3$, $\sigma(2)=1$, and $\sigma(3) =2$.
The iterative application of the resulting permutation ($\sigma$ creates the following loop: $1 \to 3 \to 2 \to 1 \to \ldots$.
Then $C(\sigma) = 1$.
In another example, $\sigma_i(1) = 2$, $\sigma_i(2) =1$, $\sigma_i(3) =3$ $\sigma_j(1) = 2$, $\sigma_j(2)=1$, and $\sigma_j(3) =3$.
Then, $\sigma(1) = 1$, $\sigma(2)=2$, and $\sigma(3) =3$.
The resulting permutation $\sigma$ becomes the identity $e$ and creates three loops, namely $1 \to 1$, $2 \to 2$, and $3 \to 3$.
Then $C(\sigma) = 3$.

The permutation model helps analyze the entanglement entropy of the random tensor networks \cite{Romain2019} and random quantum circuits \cite{Yimu2020}.
Although the present study did not focus on deriving the permutation model, we have provided a brief introduction for readers.
A random tensor network model was proposed in the context of a recent study on entanglement entropy to analytically investigate a wide range of cases.
We considered a structured bulk network of locally connected tensors randomly drawn from a uniform distribution to produce an ensemble of random tensor network (RTN) states.
Averaging random tensors effectively removed all details in the many-body state, leaving only the entanglement features encoded in the network structure and its dimensionality.
In particular, the replica method enabled us to explore the physics of random tensor networks at arbitrary bond dimensions ($D_e$, which characterize the Hilbert space of a system, exploring both the volume and area-law states.
A critical point exists between the volume and area-law states.
However, its derivation is nontrivial.
Moreover, classical numerical computations are inefficient for this purpose.
Therefore, an analytical method that can yield precise results is desirable.

A random tensor network was constructed as follows.
We prepared a bulk system on a graph $G=(V,E)$ and set the basis state labeled as $\mu =1,2,\ldots, D_e$ of the bulk system on each bond $e \in E$.
A random tensor was set on each vertex $v \in V$.
Here, we considered the graph to be a square lattice.
R\'{e}yni measures the entanglement entropy.
We employed the replica method to average the random tensor and computed the following quantities.
\begin{eqnarray}
    S^{(m)}_A = \frac{1}{1-m} \lim_{n \to 0}\frac{1}{n}\left[\left({\rm Tr}(\rho^m_A)\right)^n - \left({\rm Tr}(\rho^m)\right)^n\right]_{\rm RTN}
\end{eqnarray}
Here, $\rho_A$ and $\rho$ are the density matrices of the subsystem $A$ and the entire system, respectively.
The random tensor network model considers the boundary as a subsystem.
The exponent $m$ is the R\'{e}nyi entropy, and the index $n$ is the replica method.
The brackets with RTNs denote the configurational average over random tensors drawn from an independent, identically distributed Gaussian distribution.

The effective model of the random tensor network is given by Equation (\ref{rtn}), and $J=\log D_e$.
Thus, the critical point of the random tensor network model is characterized by $D_e$.
An analysis of the infinite bond dimension $D_e \to \infty$ describes the physical properties
of the RTNs above the critical point \cite{Hayden2016}.
However, a nontrivial state exists in bond dimensions lower than the critical point, and it qualitatively differs from the case with a higher bond dimension, corresponding to the breakdown of the Ryu-Takayanagi
formula \cite{Ryu2006}.

{\it Duality analysis of symmetric group:}
We generalize the duality analysis to the symmetric group $S_q$.
We consider the following partition function of the permutation model.
\begin{eqnarray}
    Z = \sum_{\{\sigma\}} \prod_{(ij) \in E} f(\sigma_i \sigma_j^{-1}).
\end{eqnarray}
Here, $(ij) \in E$ is an edge on the square lattice, and $\sigma_i$ denotes the permutation at each site.
The degrees of freedom are expressed by $q!$ permutation.
The $q=2$ case recovers the standard Ising model.
We use $\rho^{\lambda}(\sigma)$ to represent the symmetric group $S_q$, a matrix-valued function. Here, $\lambda$ is a partition of an integer $q$. 
The partition of a positive integer $q$ is a sequence of positive integers,
$\lambda = (\lambda_1,\lambda_2,\ldots,\lambda_l)$ satisfying $\lambda_1 \ge \lambda_2 \ge \ldots \ge \lambda_l>0$ and $q=\lambda_1+\lambda_2,\ldots,\lambda_l$.
We write $\lambda \vdash q$ to denote that $\lambda$ is a partition of $q$.
For instance, the number $4$ has five partitions: $(4),(3, 1),(2, 2),(2, 1, 1),(1, 1, 1, 1)$. 
We can also pictorially represent partitions using Young's diagrams.
A Young tableau was obtained by filling in the boxes of a Young's diagram with numbers.
The number of young tableaus is denoted as $d_{\lambda}$.
This representation satisfies $\rho^{\lambda}(\sigma_1)\rho^{\lambda}(\sigma_2) = \rho^{\lambda}(\sigma_1 \sigma_2)$ for the same partition.
As the representation is homomorphic, $\rho^{\lambda}(g^{-1}) = \rho^{\lambda}(g)^{-1}$.
Additionally, $\rho^{\lambda}(g)^{-1} = \rho^{\lambda}(g)^{T}$.
That is, $\rho^{\lambda}(g)$ is an orthogonal matrix.
This is analogous to the property $\exp(i(\theta_1+\theta_2)) = \exp(i \theta_1)\exp(i \theta_2)$ for a conventional sinusoidal basis.
In addition, the elements of the matrix $\rho^{\lambda}(\sigma)$ satisfy the following orthogonal relation:
\begin{equation}
    \frac{1}{q!} \sum_{\sigma \in S_q}  \rho^{\lambda}_{ij}(\sigma)\rho^{\lambda'}_{kl}(\sigma) = \frac{1}{d_{\lambda}}\delta_{ik}\delta_{jl}\delta_{\lambda \lambda'}.
\end{equation}
We write the Kronecker delta as , and it takes units only when the partitions are isomorphic $\lambda \cong \lambda'$.
We define the character of the matrix representation $\rho$ as $\chi^{\lambda}(g) = {\rm tr}(\rho^{\lambda}(g))$.
This character satisfies the following orthogonality:
\begin{eqnarray}
    \frac{c(\sigma)}{q!}\sum_{\lambda \vdash q}\chi^{\lambda}(\sigma)\chi^{\lambda}(\sigma') = \delta(\sigma,\sigma'),
\end{eqnarray}
Here, the Kronecker delta takes units only when the permutation is in the conjugacy class $\sigma = \tau \sigma' \tau^{-1}$ where $\tau \in S_q$.
The number of conjugacy classes is denoted as $c(\sigma)$.
The trivially for the identity $e$, $c(e) = 1$. 
We use one of the above identities as follows:
\begin{eqnarray}
    \delta\left(e, \prod_{(ij) \in \square_k} \sigma_{ij} \right)  = \frac{c(e)}{q!}\sum_{\lambda \vdash q} d_{\lambda} \chi^{\lambda}\left(\prod_{(ij) \in \square_k} \sigma_{ij}\right),\label{Eqdelta}
\end{eqnarray}
where $\chi^{\lambda}(e) = d_{\lambda}$.
We perform the summation of the over vertex spins $\sigma_i$, which are replaced by the summation over bond spin $\sigma_{ij}$, with the insertion of the products of the delta functions introduced in Equation (\ref{Eqdelta}).
The character function is reduced to a product over $(ij) \in \square_k$, as follows:
\begin{eqnarray}
   \chi^{\lambda}\left(\prod_{(ij) \in \square_k} \sigma_{ij}\right) = {\rm Tr}\left( \prod_{(ij) \in \square_k} \rho^{\lambda}\left( \sigma_{ij}\right) \right).
\end{eqnarray}
We explicitly write the indices of the matrices as
\begin{eqnarray}
 {\rm Tr}\left( \prod_{(ij) \in \square_k} \rho^{\lambda}\left( \sigma_{ij}\right) \right)
 = \sum_{\{ l_i \}} \rho_{l_1,l_2}^{\lambda_k}(\sigma_{12})\rho_{l_2,l_3}^{\lambda_k}(\sigma_{23})\rho_{l_3,l_4}^{\lambda_k}(\sigma_{34})\rho_{l_4,l_1}^{\lambda_k}(\sigma_{41}).
\end{eqnarray}
The index $l_i$ of the matrix representation of $\rho^{\lambda}(\sigma_{ij})$ emerges at each site $i$ in a square lattice.
The trace ensures that subscripts that appear twice have the same index.
We obtain the following interaction term in the partition function between adjacent plaquettes $(k,k')$:
\begin{eqnarray}
\sum_{\sigma_{ij}}f(\sigma_{ij}) \tilde{\rho}_{l_{i},l_{j}}^{\lambda_k}(\sigma_{ij}) \tilde{\rho}_{l'_{j},l'_{i}}^{\lambda_{k'}}(\sigma_{ji})  = \left[ 
 \hat{f}\left( \lambda_k, \lambda_{k'} \right) \right]_{(l_i,l'_i),(l_j,l'_j)},\label{duality}
\end{eqnarray}
where $\tilde{\rho}^{\lambda}_{ij}(\sigma) =(d^\lambda/q!)^{1/4}\rho^{\lambda}_{ij}(\sigma)$).
We reach the dual permutation model in the symmetric group.
\begin{eqnarray}
    Z = q! \sum_{\left\{ \lambda \right\}} \prod_{(kk') \in E} \hat{f}(\lambda_{k},\lambda_{k'}).
\end{eqnarray}
The dual model has degrees of freedom for the partitions of integer $q$.
Thus, the permutation model is not self-dual.
The generalization of the Kramers and Wannier duality is discussed in \cite{Buchstaber2003}.
The duality relationship is also given for the interaction strength of the integer $q$.
However, rewriting the partition function using the character function has not yet been detailed.
This is highly nontrivial, and we offer an explicit transformation of the partition function.
Furthermore, the dual model is ambiguous.
An orthogonal basis can be given by both the character function ($\chi_{\lambda}(\sigma)$) and representation ($\rho_{\lambda}(\sigma$)) of the symmetric group.
In this sense, a generalized Fourier transform can be obtained in two ways.
In the present study, we use a representation to transform the partition function and obtain the duality relation, as in Equation (\ref{duality}).

Following the replica method in the spin-glass theory, we consider the analytical continuation of $q$ into a real value to obtain the limit of $q \to 0$.
Therefore, we set a simple equality to estimate the critical points.
Here, we conjecture that the following equality estimates the location of the critical points, similar to other spin models: 
\begin{eqnarray}
    f(e) = \hat{f}((q),(q)), 
\end{eqnarray}
where $(q)$ is the trivial partition of integer $q$.

We apply this equation to the random tensor network model and obtain the following equation:
\begin{eqnarray}
    D_e^{mn} = \frac{1}{\sqrt{\Gamma(mn+1)}}\frac{\Gamma(D_e+mn)}{\Gamma(D_e)}.\label{rep}
\end{eqnarray}
where we use $q=mn$ and the identity $\sum_{\sigma \in S_{q}} x^{C(\sigma)} = \Gamma(q+x)/\Gamma(x)$.
The trivial case $mn=2$ recovers the case of the Ising model as $\exp(-J) = \sqrt{2}-1$.
Notably, the energy scale is half that of the standard Ising model.
The second simplest case, $mn=3$ leads to 
\begin{equation}
    1 = \frac{1}{\sqrt{6}}\left(1+\frac{1}{D_e}\right)\left(1+\frac{2}{D_e}\right).
\end{equation}
The same equation can be obtained by generalizing the Kramers and Wannier duality $(f(e)+3f(g_1)+2f(g_2))/\sqrt{6}f(e)$, where $f(e)=\exp(3J)$, $f(g_1)=\exp(J)$ and $f(g_2)=\exp(2J)$ as in a previous study \cite{Buchstaber2003}.

We take the limit of $n \to 0$ in Equation (\ref{rep}) and obtain
\begin{eqnarray}
    \log D_e = \frac{\gamma}{2} + \Psi(D_e).
\end{eqnarray}
where $\gamma$ is the Euler-Mascheroni constant, and $\Psi(x)$ is the so-called digamma function.
The critical point of the random tensor network model on a square lattice is $D_e = 1.882$.
The critical point value of the random tensor network model is unknown.
However, this is consistent with the natural expectation that the critical point is in the low-bond dimensions ($1<D_e<2$ \cite{Romain2019}.

{\it Conclusion:}
We estimated the critical point of the random tensor network model using duality analysis in the symmetry group.
Our conjecture was reasonable through the analytical continuation in the replica method and several validations in the simple cases.
In addition, the natural expectation of the critical points of the random tensor network model was consistent with our estimation.

Future research should focus on the numerical validation of our conjecture.
Our technique is available in several effective models of random quantum circuits \cite{Yimu2020}.
We need a generalization of a star-triangle transformation to a permutation model and the duality analysis \cite{Nishimori2006,Ohzeki2006,Ohzeki2007proc,Ohzeki2007}.
Furthermore, the duality analysis is also available for bond dilution, closely related to the ``quenched'' quantum gravity.
In other words, the entanglement entropy on a random graph can be addressed via a duality analysis \cite{Nishimori1979dilution, Ohzeki2012}.
In addition, the random bond dimension ($D_e$) for each bond must be within the duality analysis range, as discussed in \cite{Romain2019}.

Furthermore, systematic improvements through a partial summation are important.
This deviation from our conjecture indicates the poor precision of our results.
In addition, the precise locations of the critical points support the numerical investigation of the critical behavior in quantum many-body systems.
Random tensor networks and quantum circuits are particularly harmful to classical numerical computations.
Current quantum computers cannot precisely estimate the location of critical points or investigate critical behaviors.
Therefore, a systematic analytical approach is important for understanding quantum many-body physics, even if the technique is poor.

We thank Ryuki Ito for identifying the first calculation of this critical point.
This study was supported by the JSPS KAKENHI Grant No. 23H01432.
Our study received financial support from the public\verb|\|private R\&D investment strategic expansion prograM (PRISM) and programs for bridging the gap between R\&D and ideal society (Society 5.0) and generating economic and social value (BRIDGE) from the Cabinet Office.

\bibliographystyle{ptephy}
\bibliography{main}
\end{document}